\documentclass[preprint]{aastex}
\received{}
\revised{}
\usepackage{amsmath}
\usepackage{amssymb}
\usepackage{graphics}
\usepackage{graphicx}
\usepackage{natbib}
\usepackage{gensymb}
\usepackage{pdflscape}
\usepackage{capt-of}
\usepackage{morefloats}
\usepackage{xcolor}
\usepackage{epsfig}
\usepackage[normalem]{ulem}
\shorttitle{Non-thermal velocities and its CLV in Active Regions}
\shortauthors{Ghosh, Tripathi and Klimchuk}
\begin{document}
\title{Non-thermal Velocity in the Transition Region of Active Regions and its Centre-to-Limb Variation}
\author{Avyarthana Ghosh\altaffilmark{1}, Durgesh Tripathi\altaffilmark{1}, James A. Klimchuk\altaffilmark{2}}
\affil{$^1$Inter-University Centre for Astronomy and Astrophysics, Post Bag - 4, Ganeshkhind, Pune 411007, India}
\affil{$^2$NASA Goddard Space Flight Center, Code 671, Greenbelt, MD 20771, USA}
\date{}
\begin{abstract}

We derive the non-thermal velocities (NTVs) in the transition region of an active region using the \ion{Si}{4}~1393.78~{\AA} line observed by the Interface Region Imaging Spectrograph (IRIS) and compare them with the line-of-sight photospheric magnetic fields obtained by the Helioseismic and Magnetic Imager (HMI) onboard the Solar Dynamics Observatory (SDO). The active region consists of two strong field regions with opposite polarity, separated by a weak field corridor, that widened as the active region evolved. The means of the NTV distributions in strong-field regions (weak field corridors) range between $\sim$18{--}20 (16{--}18)~km~s$^{-1}$, albeit the NTV maps show much larger range. In addition, we identify a narrow lane in the middle of the corridor with significantly reduced NTV. The NTVs do not show a strong center-to-limb variation, albeit somewhat larger values near the disk center. The NTVs are well correlated with redshifts as well as line intensities. The results obtained here and those presented in our companion paper on Doppler shifts suggest two populations of plasma in the active region emitting in \ion{Si}{4}. The first population exists in the strong field regions and extends partway into the weak field corridor between them. We attribute this plasma to spicules heated to $\sim$0.1 MK (often called type II spicules). They have a range of inclinations relative to vertical. The second population exists in the center of the corridor, is relatively faint, and has smaller velocities, likely horizontal. These results provide further insights into the heating of the transition region. 

\end{abstract}


\section{Introduction}

An image of the Sun recorded at coronal temperatures consists of three distinct features, namely active regions (ARs), coronal holes and quiet Sun. Out of which, the ARs consist of the strongest magnetic fields and demonstrate the most pronounced heating. Besides, it is observed that most of the flares and coronal mass ejections originate from active regions \citep{TriBC_2004, WebH_2012, Ben_2017}. Therefore, it is mandatory to fully understand the local as well as global structure of plasma motions in ARs and compare those with the structure of the magnetic fields. 

We have performed a detailed study of Doppler shifts in the transition region of active regions and its center-to-limb variation (CLV) in \cite{GhoKT_2019}; hereafter referred as Paper~\rm{I}. Here, we present the study of non-thermal velocities (NTVs). The NTVs are, generally, spatially unresolved and estimated by studying the excess width (beyond the thermal and instrumental widths) of the spectral lines. To have a comprehensive understanding of plasma flows in the solar atmosphere, the Doppler shift and non-thermal velocities are two equally important physical parameters. While Doppler measurements provide an idea of bulk motion of the plasma, non-thermal width can give information on unresolved motions. 

It is believed that the excess width in the spectral lines could be caused by a variety of phenomena, e.g. draining flows associated with nanoflares \citep{Par_1988, PatK_2006}, waves \citep{BolDF_1975, CheDF_1979, MarFD_1978, McCHA_1991, DerM_1993} and turbulence \citep{Jor_1980, Car_1996, InnBG_1997, SarSE_1997} etc. Therefore, an accurate measurement of this excess width may provide crucial information on the heating of the active regions in transition region and corona. 

Over the last several decades a lot of observations have been dedicated to the study of non-thermal plasma flows in active regions formed in the transition region. Some of the earliest observations have been recorded by the Orbiting Solar Observatory \citep[OSO-8;][]{Bru_1977}, the Naval Research Laboratory (NRL) normal incidence spectrograph on Skylab (S082-B) \& High Resolution Telescope and Spectrograph \citep[HRTS;][]{BarB_1975} and the Ultraviolet Spectrometer and Polarimeter \citep[UVSP;][]{WooTB_1980} onboard the Solar Maximum Mission \citep[SMM;][]{Sim_1981}. These observations have shown that there is a significant amount of non-thermal flows, ranging between 10{--}30~km~s$^{-1}$, in the transition region for solar spectral lines formed below 2000~{\AA} \citep{McAW_1972, BruM_1972, BruPC_1973, KohPR_1973, BolEJ_1973, ChiB_1975}. Authors $viz.,$ \cite{CheDF_1979} attributed this broadening to Alfv\'en wave propagation, for transition region and coronal lines. 

For an insight to the role of magnetic field structures for non-thermal flows, CLV studies involving optically thin spectral lines (\ion{O}{1}, \ion{S}{1}, \ion{N}{5}, \ion{O}{4}, \ion{C}{4} and \ion{Si}{4}) have been conducted by several authors in different solar features using observations from Skylab and OSO-8 \citep[see e.g.,][]{ShiRB_1976, MarFD_1978}. \cite{FelDP_1976} reported that the resonance lines of \ion{C}{4} and \ion{Si}{4} have an increased line width at the limb, attributed to enhanced opacity unlike the optically thin lines $viz.,$ \ion{O}{1}, \ion{O}{4}, \ion{N}{5} $etc$. Likewise, \cite{RouFB_1979} reported an equivalent excess broadening of 3~km~s$^{-1}$ at the limb as compared to the disk center for \ion{Si}{4} 1393~{\AA} line. Similarly, \cite{DoyW_1980} showed that transition region lines such as \ion{Si}{4} 1393~{\AA} and \ion{C}{3} lines have predominant disk-to-limb enhancement, again with opacity being the plausible cause. On the contrary, \cite{AthW_1980} reported nearly constant line widths for \ion{C}{4} 1548~{\AA} as a function of disk position, except at far-limb regions \citep[also see,][]{KjeN_1977, DerBB_1984}. 

\cite{Kli_1985} performed an extensive study over a number of features comprising of ARs and sunspots with \ion{C}{4} line 1548~{\AA} ($\log\, T=5.05$) observations from the UVSP. The author reported that for active regions, except sunspots, the Doppler width is $\sim$0.151~{\AA} \citep[also see,][]{AthGH_1983a}, which translates to 30~km~s$^{-1}$, without any correction for instrumental broadening. For the sunspots, however, it reduces to 0.121~{\AA} ($\sim$23~km~s$^{-1}$), in contradiction to the results obtained by \cite{GurLS_1982}, typically about 0.097~{\AA} (19~km~s$^{-1}$) for centers of sunspot umbrae. Moreover, the variation of the linewidths, as reported by the author, does not reveal convincing CLV \citep[also refer to][]{AthW_1980}. It was further noted that the line widths were larger for redshifted features as compared to the blueshifted ones. 

Subsequently, \cite{DerBB_1987} used quiet Sun as well as active region observations from the Spacelab-2 on HRTS in the \ion{C}{4} line at different locations on the disk. The author reported that the average Full Width Half Maxima (FWHM) is $\sim$0.195~{\AA} with a most-probable non-thermal velocity of 19~km~s$^{-1}$. Further, authors, viz., \cite{HasRS_1990, DerM_1993} reported that the transition region lines have non-thermal velocities 24{--}32~km~s$^{-1}$ in active regions \citep[see also,][]{MarFD_1979, HasRS_1990}.

In more recent times, \cite{ErdDP_1998} concluded that the non-thermal velocities increase significantly at the limb as compared to the disk for quiet Sun spectral lines in both chromospheric and coronal lines such as \ion{C}{1}, \ion{Fe}{2}, \ion{S}{4}, \ion{O}{6}, \ion{Mg}{10}, \ion{Fe}{12} observed by the Solar Ultraviolet Measurements of Emitted Radiation \citep[SUMER;][]{WilCM_1995} on-board the SOlar and Heliospheric Observatory \citep[SOHO;][]{DomFP_1995}. This increase was attributed to opacity effects. 

In addition to these estimates, a positive correlation between non-thermal widths and intensities of features, on a spatially larger scale, viz., as active regions or quiet Sun features being treated as a whole, has also been reported by several authors, $viz.,$ \cite{AthGH_1983a, DerBB_1984, BruM_1979, Kli_1985}. Based on SUMER observations of a variety of solar structures, \cite{Fel_2011} observed that in ARs, the pixels with higher intensities also had higher excess line widths and predominant redshifts \citep[also see,][]{GebHN_1981, Feld_1983}.

With EIS observations, studies have been done to augment our understanding of non-thermal velocities in various structures in active regions such as loops and moss \citep{DosMW_2007, Del_2008, BroW_2009}. However, the transition region lines observed with EIS ($viz.,$ \ion{O}{4}, \ion{O}{5} and \ion{Mg}{5}) are very weak \citep{YouDM_2007} and the studies resulted in very few conclusive results.

With the very high spatial resolution observations of \ion{Si}{4} recorded by the Interface Region Imaging Spectrograph \citep[IRIS][]{DePTL_2014a}, \cite{DePMM_2015} showed that the non-thermal flows in active regions remain constant ($\sim$20~km~s$^{-1}$), independent of position on the disk. Similar results were deduced for the quiet Sun and coronal holes as well. To explain these, the authors suggested combined effects of shocks ($\parallel B$; parallel to magnetic field) and twisting motions ($\perp B$; perpendicular to magnetic field). However, we note that \cite{DePMM_2015} studied the NTV averaged over the entire active regions. 

In the present study, we conduct a study of non-thermal velocities and its CLV in an AR, by dividing the ARs into strong field regions and weak field corridors running in between the strong field regions. Therefore, it should be possible to seek out any differences in the non-thermal velocities in these regions, based on the line-of-sight (LOS) magnetic flux densities. We emphasise that this is the first time such a study, i.e. dividing the active region in strong field and weak field corridor, is being done with any modern spectrograph.

As mentioned earlier, in Paper~\rm{I} we have discussed the Doppler shifts in the transition regions of active regions and its CLV. In this paper, we continue our study of non-thermal velocities with observations of the same active region, viz., \textsl{AR~12641} from the IRIS and the LOS magnetic field observations from the Helioseismic Magnetic Imager \citep[HMI;][]{SchBN_2012, SchSB_2012} onboard the Solar Dynamic Observatory \citep[SDO;][]{PesTC_2012}. The rest of the paper is structured as follows. In \S\ref{obs}, we provide a detailed description of the data used, a brief description of the instruments used and discuss the processing techniques. In \S\ref{analy}, we present the method of analysis and results. We summarise the results and discuss those in \S\ref{sum}. Finally we conclude in \S\ref{conc}.

\section{Observations} \label{obs}

Since the main aim of this work is to study the non-thermal width and its CLV in an evolving active region, we considered the same rasters of active region \textsl{AR~12641}, which we have used to study the Doppler shifts and its CLV in active region in \textsl{Paper \rm{I}}. This active region was observed with IRIS using 320-step dense raster, with each step having an exposure time of 4~s between February 28 and March 9, 2017 while it crossed the central meridian. As mentioned in \textsl{Paper \rm{I}}, the \ion{Fe}{2} line used for wavelength calibration was not well-defined on February 28. Also, on March 9, the AR under consideration was very close to the west limb. To maintain parity between the analyses done in \textsl{Paper \rm{I}} , we have excluded the observations taken on February 28 and March 9 here as well. 

\begin{figure}[t!] 
\centering
\includegraphics[width= 1.0\textwidth]{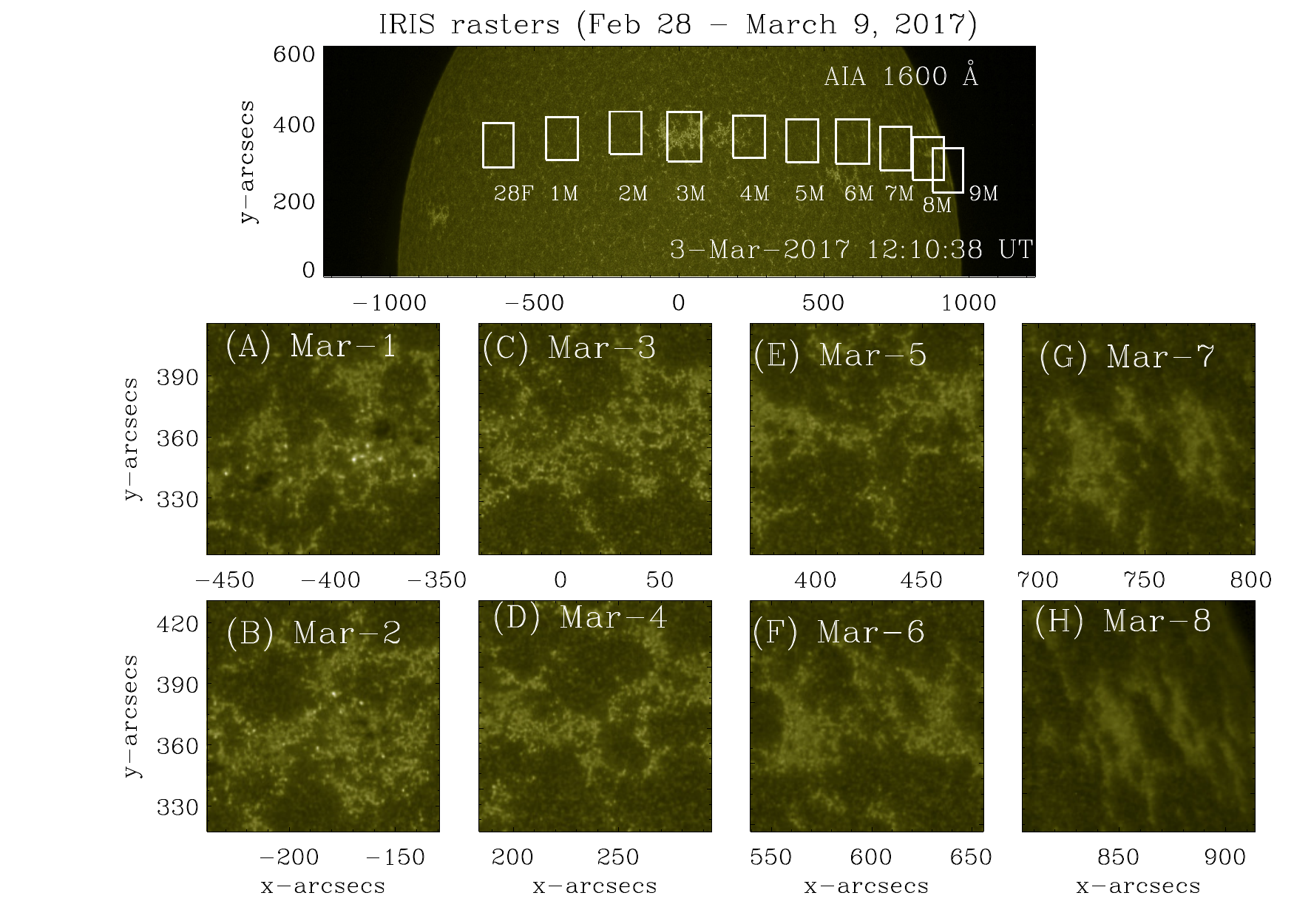}
\caption{Top panel: Partial disk image of the Sun taken in the 1600~{\AA} channel of AIA on March 3, 2017 (also see Paper-I). The over-plotted white-boxes locate the IRIS raster field of view from 28th Feb until March 9th of active region \textsl{AR 12641} as it passed through the central meridian. The letter `F' stands for February and `M' for March. Middle and bottom rows: Zoomed-in FOVs of AIA 1600~{\AA} channel between March 1{--}8, 2017.}\label{track}
\end{figure}

In the top panel of Fig.~\ref{track}, we display the IRIS fields-of-view (FOV) on different days as labelled, over-plotted on portion of full disk of the Sun recorded by AIA in the 1600~{\AA} channel on March~3, when the active region was located at the central meridian. In the first and second columns of Table~\ref{tab_ntv}, we provide the date and $\mu$-value of each observation, where $\mu$ is defined as the cosine of the angle between the surface normal and the LOS to the observer of the centre of the FOV \citep{Tho_2006}\footnote{We note the definition of $\mu$ written in \textsl{Paper \rm{I}} is not complete. This is the definition we have used in \textsl{Paper \rm{I}}.}
 
We have used Level-2 IRIS raster data in which all instrumental effects such as flat-fielding, dark currents, offsets and thermal orbital variations have been accounted for, so as to make it suitable for scientific analysis\footnote{A User’s Guide To IRIS Data Retrieval, Reduction $\&$ Analysis, S.W. McIntosh, February 2014}. Further, we apply the procedure called the Intensity Conserving Spectral Fitting \citep[ICSF;][]{KliPT_2016} on the original line profiles, as explained in \textsl{Paper \rm{I}}. We use single Gaussian fitting routines\footnote{Using EIS Gaussian fitting routines for IRIS data, P. Young, April 2014} on the resultant spectral lines, provided in \textsl{Solarsoft} \citep{FreH_1998} for analysing the data. Since our aim is to measure the non-thermal velocities, we choose the stronger \ion{Si}{4} line at 1393.78~{\AA} as compared to the \ion{Si}{4} line at 1402.77~{\AA} \citep[see e.g.,][]{GonV_2018, TriNI_2020}.

\begin{figure}[t!] 
\centering
\includegraphics[width= 1.0\textwidth]{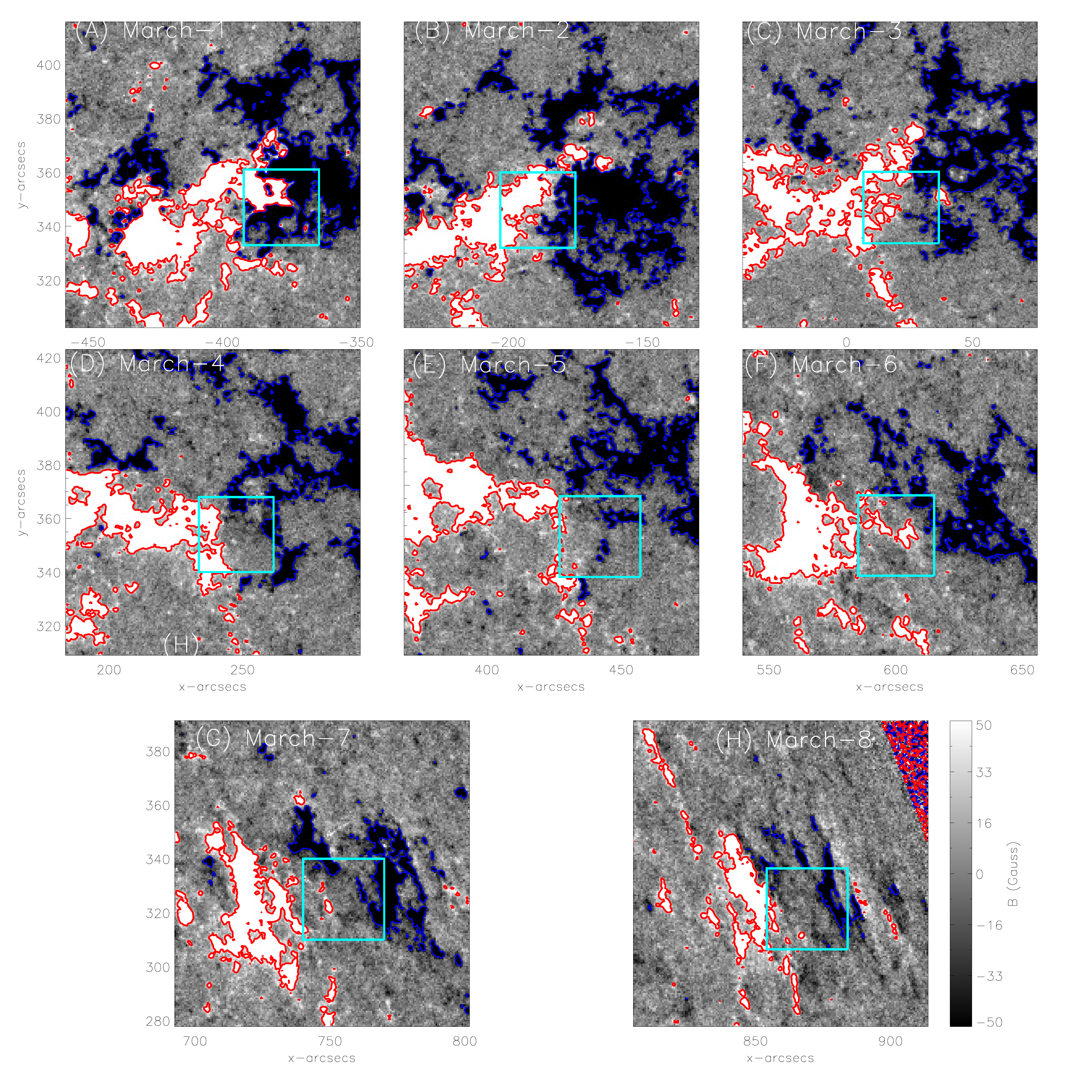}
\caption{Photospheric LOS magnetograms from HMI of the AR under study as it passes  across the central meridian. The over-plotted red and blue contours show magnetic flux density of $\pm50$~G, respectively. The cyan colored box in each panel indicates the corridor region.}\label{hmi}
\end{figure}

Furthermore, we have used the photospheric LOS magnetograms obtained by the HMI on-board SDO to compare between non-thermal motions and the structure of magnetic fields in the photosphere. The photospheric magnetic field of active regions can be roughly divided into areas of strong and weak vertical field \citep{Kli_1987}. The gap of weak field that separates strong field regions of opposite polarity is called a corridor. We distinguish strong and weak fields in the observations using a threshold of 50~G in longitudinal (LOS) magnetograms. Because the LOS component of a vertical field diminishes away from disk center, the 50~G threshold only approximately identifies the boundary between strong and weak vertical field. By examining different thresholds, we conclude that the consequences of this approximation are only modest over much of the disk but can be substantial close to the limb. Consequently, the statistics we present for observations near the limb have some ambiguity and are less significant. In Fig.~\ref{hmi} we plot the corresponding LOS photospheric magnetograms for the active region as it rotated across the disk. The red and blue contours represent contours of magnetic flux density of $\pm$50~G, respectively. As can be seen, the threshold of $\pm$50~G isolate the strong field regions and weak field corridor very well. The superimposed cyan box in each panel highlight the corridor region separating the strong field regions on respective dates.

For the process of co-alignment between IRIS and HMI observations, we have followed the same scheme as described in \textsl{Paper \rm{I}}. In brief, we have first co-aligned HMI magnetogram to AIA image taken at 1600~{\AA}. The IRIS \ion{Si}{4} intensity images were then co-aligned with those of AIA 1600~{\AA}. 

\section{Data Analysis and Results} \label{analy}
\subsection{Non-thermal Velocities}\label{sec_ntv}

We measure the broadening of a spectral line by its FWHM, given as

\[
\mathit{FWHM}= (\lambda_{0}/c)\sqrt { 4~ln 2 * \huge\big[\big(\frac{2k_{B}T}{M}\big)+v_{nth}{^2}\big] + \epsilon{^2}}
\]

\noindent where T is the temperature of the plasma, M is the mass of the ion, $\lambda_{0}$ is the rest wavelength, c is the speed of light. Here, v$_{nth}$ is the non-thermal velocity (NTV) and $\epsilon$ is the instrumental broadening, characteristic of the instrument used. The first component is thermal broadening because the ions obey the Maxwell-Boltzmann distribution of velocities and given by $(\lambda_{0}/c)\sqrt{(4 ln 2)*(\frac{2k_{B}T}{M})}$. We assume that the plasma is at $\log\,T[K]$= 4.90 and the instrumental broadening for IRIS is equivalent to 4~km~s$^{-1}$  \citep[$\sim$0.02~{\AA},][]{DePTL_2014a}.

The intensity (panels A, B, C \& G) and NTV maps (panels D, E, F \& H) obtained for March 1{--}8 observations are shown in Figs.~\ref{ntv1} and~\ref{ntv2}. In the NTV maps, the black pixels represent a combination of missing pixels (no measurement) and those where the measured FWHM is less than the thermal component to avoid imaginary values of NTV in the above equation. The over-plotted (black) contours represent magnetic flux density of 50 (-50)~G, respectively, obtained from the HMI magnetograms. These maps reveal that the strong field regions have non-thermal velocities in the range $\approx$~5{--}30~km~s$^{-1}$, irrespective of the location on the disk. The NTV maps also reveal a clear distinction between the strong field regions and the weak field corridor in the center of the active region, with values somewhat smaller than those in the strong field regions.

We plot the distribution of NTVs for strong field regions (red) and weak field corridors (blue) in Fig.~\ref{hist_ntv}. The means and standard deviations for each distribution are also labelled on respective panels. For the strong field regions (red histograms), the peak of the distributions are maintained at a constant values of  $\approx$~20~km~s$^{-1}$, with the mean values ranging between $\approx$~18{--}20~km~s$^{-1}$. However, we note that the distributions are more symmetric about the peak for disk center positions (e.g. panels C and D), whereas there is an enhanced wing on the lower velocity side as we move away from the disk center.

To precisely estimate the NTVs in the corridor, we consider regions which have been highlighted by the white boxes in each panel in Figs.~\ref{ntv1} and~\ref{ntv2}. We note that that these are the same locations which were considered for the estimation of Doppler velocities in the corridor regions in Paper I (refer to Figures 12 \& 14). By restricting ourselves to these boxes, we intentionally avoid other weak field regions, e.g., outside the active region, which do not necessarily share the same properties as the corridor (there may be many similarities, however). Fig.~\ref{hist_ntv} also displays the distribution of NTVs (in blue) obtained within the boxes. Similar to the histograms for the strong field regions, we note that the NTV distributions are also symmetric about the peak when the AR is closer to the disk center. 

Table~\ref{tab_ntv} provides the mean along with the standard errors \footnote{Refer to Appendix in Paper-I for detailed discussion} and the median values of NTVs in strong field regions (columns 3 \& 4) and those of corridors (columns 5 \& 6). We note that both the mean and median values for corridor regions are smaller than those in the strong field regions at all disk positions except for the observations taken on March 1 \& 8. Moreover, the distributions in corridor are flatter than those for the strong field regions for the locations closer to the disk center. We further note that, on March 8, when  the region is very close to the limb of the Sun, the distribution is significantly narrower than on previous dates. Moreover, the distributions on March 8 have similar number of points for strong field and weak field corridor regions. We attribute this to two things. First, the region of weak vertical field that we call the corridor is broadening with time as the active region evolves. Second, the transition between weak vertical field and strong vertical field is not abrupt, so the 50~G LOS threshold we use to differentiate the two in the observations is shifted in a manner that artificially broadens the corridor near the limb. It is likely that many of the pixels contributing to the corridor averages on March 8 are actually from regions of strong vertical field. This could explain why the averages are similar to the strong field values on that day. We, therefore, give them less significance.

\subsection{Centre-to-limb variation of Non-thermal Velocities}\label{clv_ntv}

We have further studied the CLV of NTVs in strong field regions (panel A) as well as corridors (panel B). For that purpose, we plot the mean values of NTVs as a function of {$\mu$} in Fig.~\ref{theta}, where $\mu$ closer to 1.0 (0.0) corresponds to the region located at the disk center (limb). The average non-thermal velocities range between 18{--}20~km~s$^{-1}$ (16{--}20~km~s$^{-1}$) for the strong field (weak field corridor) regions. The error bars are the standard errors that range between 0.05{--}0.11~km~s$^{-1}$.  As stated above, there is an additional uncertainty in the corridor value for March 8 ($\mu = 0.27$). 

\subsection{Doppler velocity Vs Full Width Half Maximum}\label{dp_ntv}
We note that all the IRIS rasters studied here were also used to measure the Doppler velocities in strong field regions as well as weak field corridors in \textsl{Paper~\rm{1}}.  Therefore, a combination of these two provides an opportunity to study the relationship between Doppler shifts and widths of the spectral lines. At this stage, it is important to re-iterate that from \textsl{Paper~\rm{1}}, we find that the strong field regions ($>$50~G) have predominate redshifts whereas the blueshifts are more common in the weak field corridor. This is true at all disk positions.

Figure~\ref{scatter_dv_wd} displays scatter plots between Doppler velocities and FWHM for strong field regions (top row) and weak field corridor covered by the boxed regions marked in Figs.~\ref{ntv1} \& ~\ref{ntv2} (bottom row). Note that the instrumental width of 0.02~{\AA} has already been subtracted. We further note that, for brevity, we have shown the scatter plots only for March 1, 3 and 7. The over-plotted red dashed-dotted lines locate the zero velocity.

The scatter plots reveal a strong positive correlation between redshifts (positive velocity) and FWHM, for both the strong field regions as well as the corridor. On the contrary, however, there is no such correlation observed between blueshifts (negative velocity) and FWHM. We note that the correlation between redshifts and FWHM is much stronger on March 1 \& 3, when the region is located near the disk centre. The correlation weakens on March 7, when the region is located near the west limb. 

We further note that the lower Doppler velocities on March 7 (panels C \& F of Fig.~\ref{scatter_dv_wd}), are consistent with the smaller and narrower histograms of Doppler velocities in Figs.~10 \& 13 of \textsl{Paper~\rm{I}}.

\subsection{Intensity Vs Full Width Half Maximum}\label{int_ntv}
Similar to Doppler velocities, we display the scatter plots for line intensity and FWHM in the top row of Fig.~\ref{scat2}. The red denotes the strong field regions whereas the blue represents the weak field corridor. In the bottom row, we also show the scatter plots after binning the data over 100 points (red diamonds for strong field regions and blue diamonds are for weak field corridor). The plots show strong relationship between the two physical quantities irrespective of the magnitude of magnetic flux densities, as well as their location the Sun's disk. However, we note that for the weak field corridor, the intensity vs FWHM curves turn-around and start decreasing beyond 0.18~{\AA}. 

\section{Summary and Discussions} \label{sum}

In this paper, we have studied the NTV and their CLV in the transition region of an active region (\textsl{AR~12641}) as it traversed across the central meridian, using the \ion{Si}{4}~1393.78~{\AA} line ($\log\, T[K] = 4.90$) observed by IRIS. Moreover, we have compared the structure of NTV with that of photospheric LOS magnetic flux density obtained from HMI onboard SDO. For this purpose, we have divided the active region into strong (weak) field regions defined as regions with magnetic flux density stronger (weaker) than $|50~G|$. The observations showed that the active region has two dominant, strong-field regions separated by an intermediate weak field corridor. The corridor region evolved and grew wider with the development of the active region. To the best of our knowledge, this is the first time such study on non-thermal velocities and their CLV is done with clear distinction between regions with different magnetic field configurations. We summarise our findings below.

The NTVs in strong-field regions, as obtained from the NTV maps, range between $\approx$~5{--}30~km~s$^{-1}$, with the mean values ranging between 18{--}20~km~s$^{-1}$. Similarly, the mean value of NTVs in the corridor range between 16{--}20~km~s$^{-1}$. Excluding March 8 for the reasons given earlier, the mean values for the corridor range between $\approx$16{--}18~km~s$^{-1}$. The mean and median values in the weak field corridors are consistently lower than the respective values in the strong field regions between March 2{--}7. We also note that the statistics for the corridor are computed over the boxed region and include some pixels near the periphery of the strong field regions that are not part of the corridor itself. Within a narrow lane in the middle of the corridor, we find much lower values of NTVs ($\approx$12~km~s$^{-1}$).

The results obtained here are in agreement with those obtained by \cite{DePMM_2015} using the IRIS \ion{Si}{4} 1403~{\AA} line. However, we note that \cite{DePMM_2015} did not discriminate between strong and weak field regions. The authors attributed the enhanced non-thermal broadening to propagating magneto-acoustic shocks originating in the chromosphere (also see references therein) or turbulence associated with these shocks. Similar non-thermal velocities have been observed by \cite{DerBB_1987} in active regions as well as quiet Sun and have been attributed to associated wave motions. Using HRTS observations, \cite{DerM_1993} reported typical non-thermal velocities in active regions to be 28$\pm$4~km~s$^{-1}$. Similarly, with HRTS observations of \ion{Si}{4} 1393~{\AA} line, \cite{DoySE_1997} reported non-thermal velocities as large as 40~km~s$^{-1}$ in quiet Sun and small-scale active regions at near-limb locations, but attributed this to the increased opacity near limb inferred as line intensity ratio of 1.5 instead of 2.0.

Within the moderate range of $\mu$-values studied here, we see that both the strong field regions and weak field corridor do not show any convincing CLV, considering that these vary by about only 2 and 4~km~s$^{-1}$, respectively. Our results are in accordance with those obtained by \cite{DePMM_2015} using IRIS observations, where the non-thermal flows measured to be nearly constant at $\sim$20~km~s$^{-1}$, irrespective of the position on the disk. These results are also in agreement with those of \cite{RouFB_1979}, which were derived using the \ion{Si}{4}~1393~{\AA} line. Our results are also similar to those obtained by \cite{DerBB_1987} using \ion{C}{4} line and \cite{NicBT_1977} using \ion{Si}{3}, which show no systematic change of line width on the disk of the Sun but a considerable variation above the limb for the optically thin UV lines \citep[see also,][]{DosFC_1977, MarFD_1978}. Likewise, \cite{FelDP_1976, DoyW_1980, ErdDP_1998} reported enhancements in the line width for observations recorded closer to the limb and thereby citing opacity as a plausible factor. 

Our study reveals that there is a strong correlation between redshifts and FWHM, i.e. the pixels with higher redshift show larger FWHM. The correlation is strongest for regions located near the disk centre and progressively decreases towards the limb. However, no such correlation is observed between blueshifts and FWHM. Similar to redshifts, a positive correlation is noted between intensity and FWHM for both strong fields regions as well as weak field corridor. However, for the weak field corridor, the intensities reach a maximum and then start decreasing at about 0.18~{\AA}. Our results of positive correlation between line width and intensity are in agreement with \cite{AthGH_1983a, DerBB_1984, BruM_1979, Mar_1992, DePMM_2015}. Similar results of enhanced emission (with stronger redshifts) and increased line widths have been reported in an AR (as well as quiet Sun, coronal holes) by \cite{Fel_2011} for chromospheric and transition region lines $viz.,$ \ion{C}{1}, \ion{Si}{2}, \ion{C}{4} etc \citep[also see,][]{WarMW_1997, HanBC_2000}. 

\section{Conclusions} \label{conc}

Non-thermal broadening of spectral lines could either be attributed to the presence of waves or to unresolved directed flows \citep[][]{BolDF_1975, CheDF_1979, MarFD_1978, MarFD_1979, DerM_1993, McCHA_1991, DoySE_1997, ZaqE_2009, DePMM_2015} . We note that a constant wave energy flux at constant pressure (e.g., waves propagating without dissipation through the transition region) have a fluctuating velocity that varies as T$^{\frac{1}{4}}$, for both Alfv\'en as well as sound waves. Thus, the line broadening would increase slightly with temperature. However, several observations show that line broadening peaks at around 0.3~MK and decreases further with temperature \citep[see e.g.,][]{ChaSL_1998}. If the Si IV emitting plasma we observe is physically connected to the corona as part of a traditional loop $i.e.,$ footpoints of the hot loops, then the line broadening is likely not due to waves. However, waves are not ruled out if the plasma is distinct from the loop transition region. 

The Doppler shifts in \textit{Paper \rm{I}} indicate the presence of two populations of flows. One population (\textit{Population \rm{I}}) of fast ($\approx$~10~km~s$^{-1}$) downflows exists in strong field regions and the outer part of the corridor, adjacent to the strong field regions, i.e., extending partway into the corridor. We associate this population with type II spicules. Another population (\textit{Population \rm{II}}) of slow ($\approx$~0~km~s$^{-1}$), approximately horizontal flows exists in the center of the corridor. Based on the results, we suggested that the \ion{Si}{4} emission in strong field regions comes primarily from type II spicules. We now argue that the line width observations, presented in this paper, are consistent with this picture.

There is a hint of two populations in the scatter plots of Figure~\ref{scatter_dv_wd}, as we now explain. If non-thermal line  broadening is due to unresolved directed flows, as opposed to waves or turbulence, then we would expect the broadening to increase with Doppler shift. A larger average velocity (given by the Doppler shift) is likely to be accompanied by a larger spread in velocity (given by the broadening), both because of the variation in velocity from one unresolved flux tube to the next and from acceleration/deceleration within a given flux tube. A positive correlation between broadening and Doppler shift is clearly present in Figure~\ref{scatter_dv_wd}. Furthermore, there is a suggestion of two overlapping clouds of data points. One cloud ranges roughly between Doppler shifts of 0 and +25~km~s$^{-1}$ and is inclined from lower-left to upper-right. This is associated with \textit{Population \rm{I}} flows. The second cloud ranges roughly between Doppler shifts of -10 and +10~km~s$^{-1}$ and is horizontal. This is associated with \textit{Population \rm{II}} flows. Though the evidence for two clouds is modest at best, it gives further credibility to the idea that \ion{Si}{4} emission in strong field regions comes primarily from type II spicules, and that the non-thermal broadening arises from the velocity dispersion of the spicules - both from one spicule to the next and from the acceleration and deceleration of individual spicules as they fall. One possible inconsistency with this interpretation is that the mean line widths in strong field regions do not exhibit a CLV decrease like the mean redshifts do (\textit{Paper \rm{I}}). 

In Paper \rm{I}, we have presented the discrepancy between observations of Doppler shifts (downflows) and the conventional idea of impulsive heating in the corona being the primary cause for the flows. In that context, our suggestion that most of the emission at $\sim$0.1~MK comes from type II spicules rectifies one of the main difficulties with the conventional picture of a thin transition region at the base of hot loops. Simulations of hot loops predict much less emission at lower transition region temperatures than is observed. Spicules can account for the large excess. 

We have argued that non-thermal line broadening in strong field regions can be explained by the dispersion in spicule velocities, but other sources of broadening, such as waves, turbulence, swaying and torsional oscillations of spicules, etc., cannot be ruled out. We have presented a simple argument for why waves are unlikely. The origin of the Doppler shifts and line broadening in the weak field corridor has yet to be explained.

\begin{acknowledgements}
We thank the referee for the detailed reading and helpful comments. This research is partly supported by the Max-Planck Partner Group of MPS at IUCAA. The work of JAK was supported by the Internal Scientist Funding Model at Goddard Space Flight Center (competitive work package program). AIA and HMI data are courtesy of SDO (NASA). IRIS is a NASA small explorer mission developed and operated by LMSAL with mission operations executed at NASA Ames Research center and major contributions to downlink communications funded by ESA and the Norwegian Space Center. 
\end{acknowledgements}
\bibliographystyle{apj}
\bibliography{references}

\begin{figure}[t!] 
\centering
\includegraphics[trim=0.cm 0.cm 0.cm 2.cm,width= 1.0\textwidth]{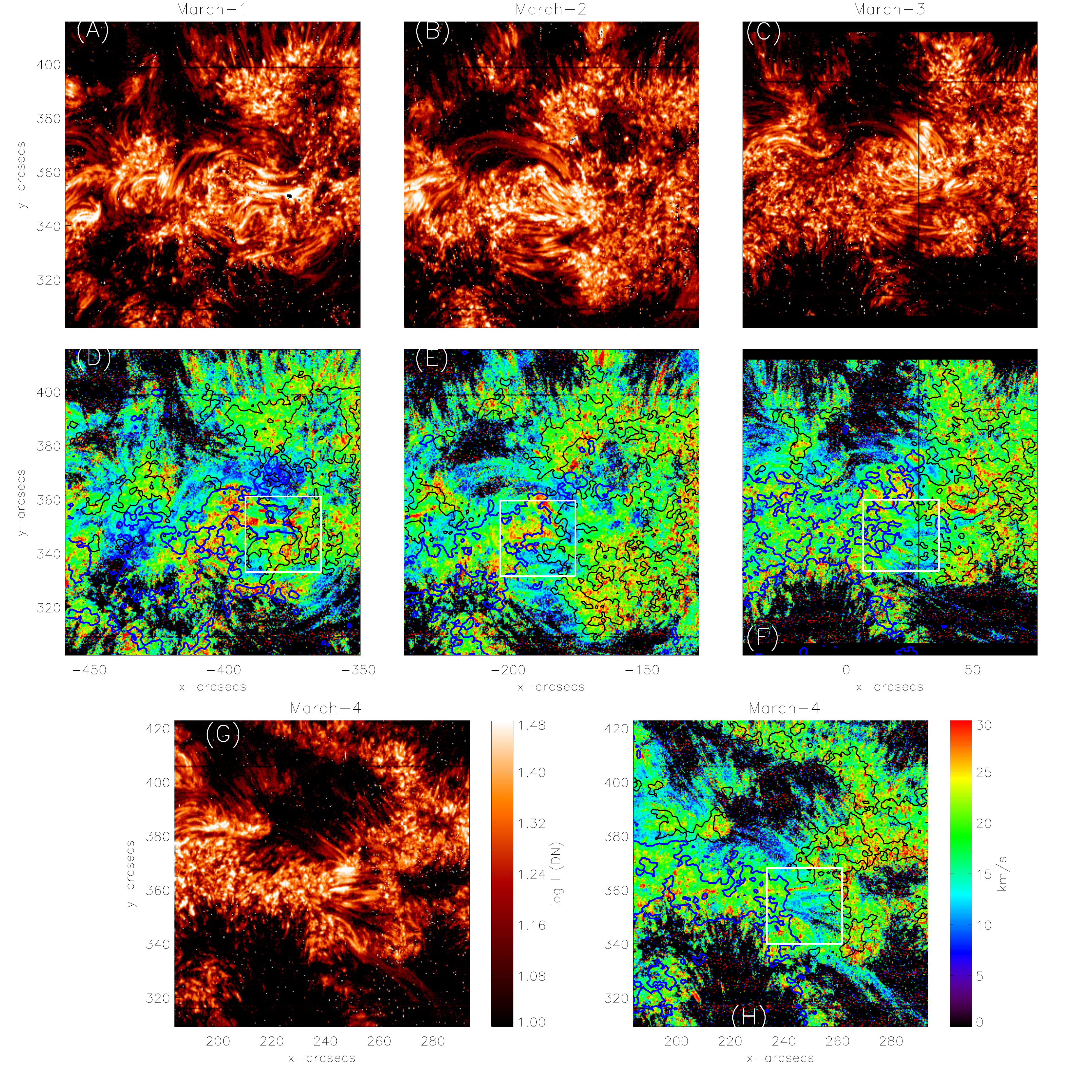}
\caption{\ion{Si}{4} 1393.78~{\AA} Intensity (panels A, B, C \& G) and non-thermal velocity (panels D, E, F \& H) maps for March 1, 2, 3 and 4. The raster on March 3 is the closest to the disk center. The overplotted white box highlights the corridor region. The blue and black contours mark the region of 50~G and -50~G, respectively.}\label{ntv1}
\end{figure}
\begin{figure}[t!] 
\centering
\includegraphics[trim=0.cm 1.cm 0.cm 4.cm,width= 1.0\textwidth]{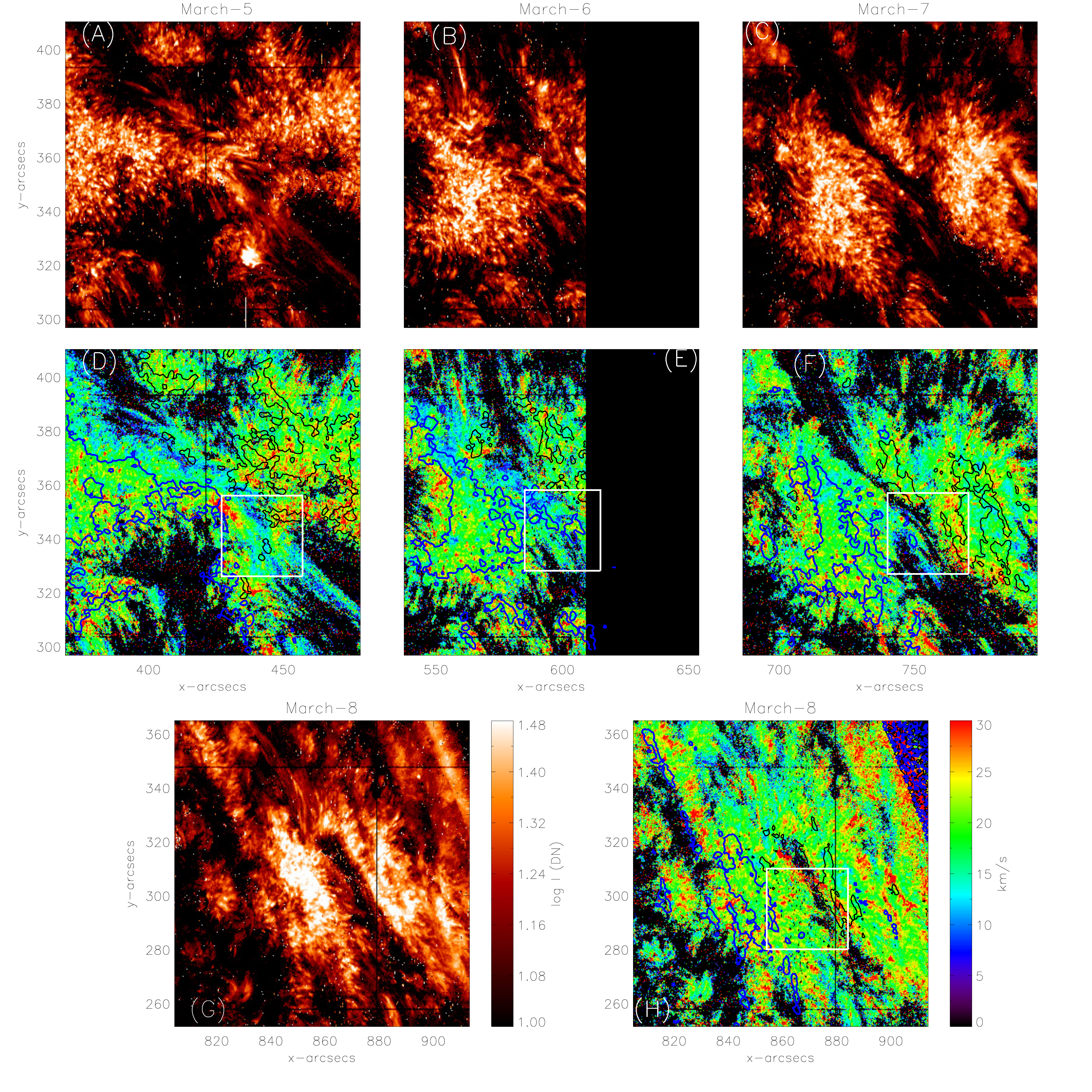}
\caption{Same as Fig.~\ref{ntv1} but for March 5, 6, 7 and 8. Note that the rasters on March 6 have a lot of missing exposures}.\label{ntv2}
\end{figure}
\begin{figure*} 
\centering
\includegraphics[trim=0.cm 2.cm 0.cm 4.cm,width= 1.0\textwidth]{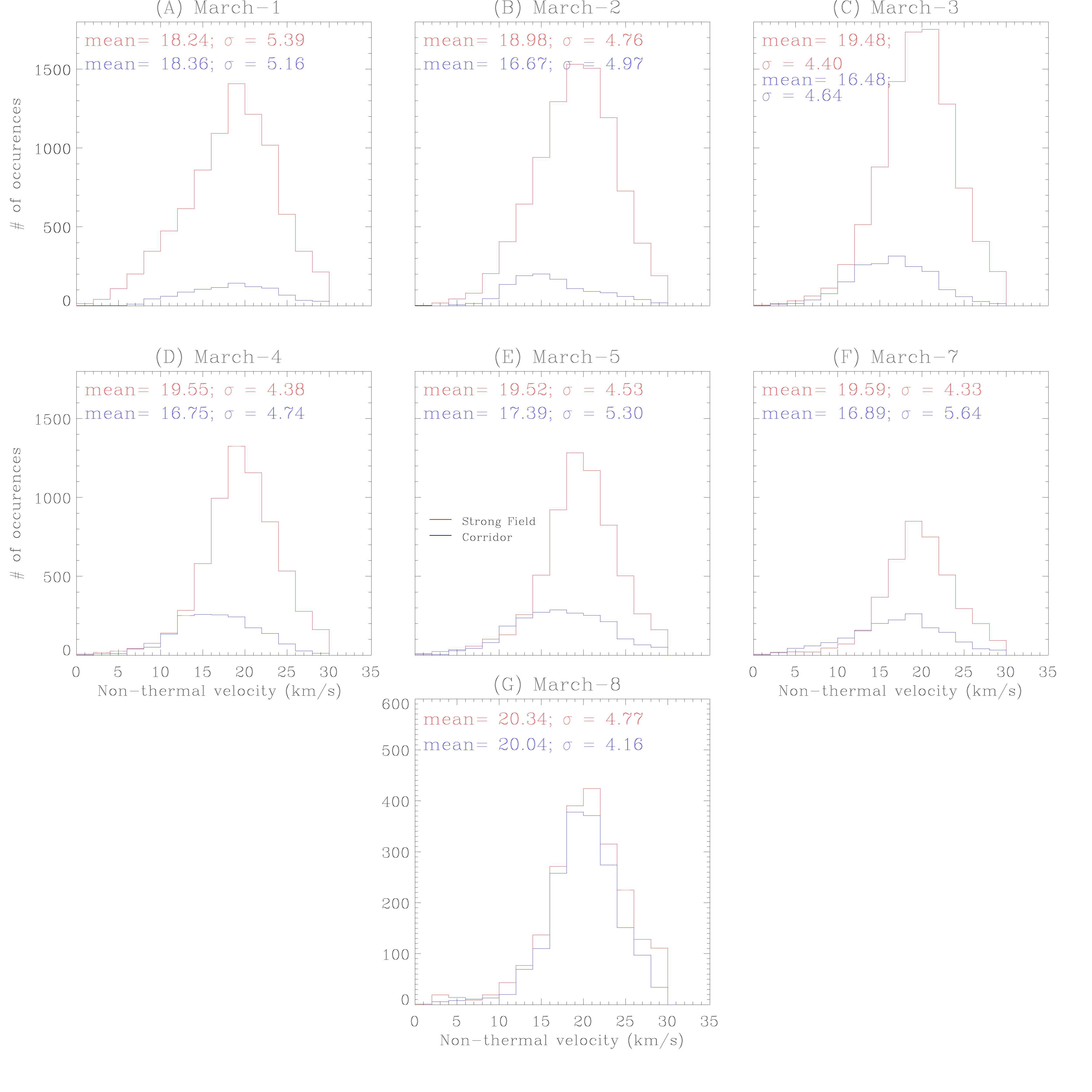}
\caption{Histograms of non-thermal velocities in the strong field regions (red) and in the corridor (blue) using \ion{Si}{4} line on the given dates. The mean and standard deviation values (in corresponding colors) are noted in each panel are in km~s$^{-1}$.}\label{hist_ntv}
\end{figure*}
\begin{figure}[t!] 
\centering
\includegraphics[trim=2.cm 5.cm 0.cm 0.cm,width=1.0\textwidth]{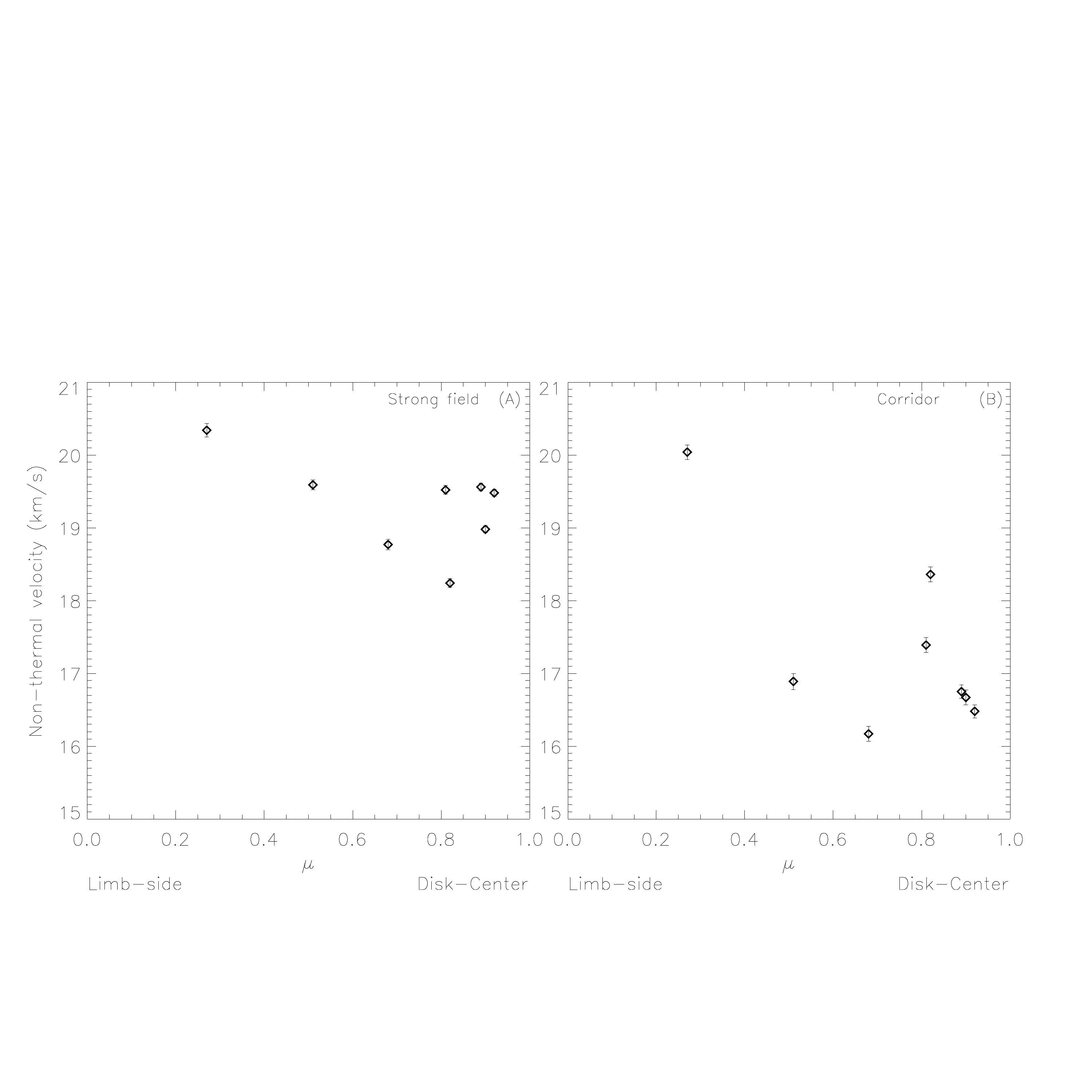}
\caption{Non-thermal velocities as a function of $\mu$ for strong field regions (panel A) and  weak field corridors (panel B). The over-plotted error bars represent the standard errors, ranging between 0.05{--}0.11~km~s$^{-1}$. Note that $\mu=$ 0.0 corresponds to closer to limb whereas $\mu$ approaching 1 is closer to disk center.}\label{theta}
\end{figure}
\begin{figure}
    \centering
    \includegraphics[trim=0cm 3cm 0cm 3cm, width=0.85\textwidth]{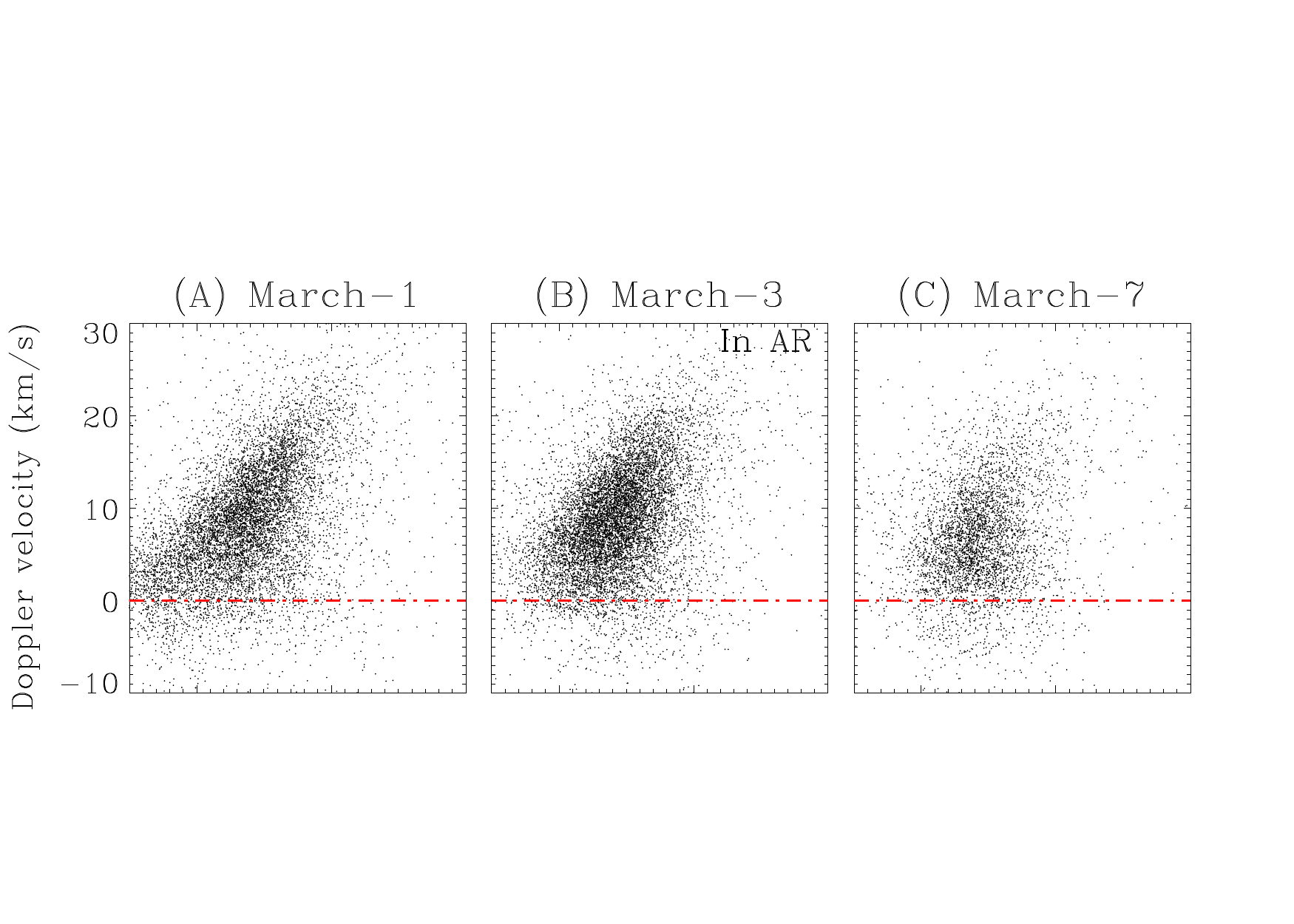}
    \includegraphics[trim=0cm 2cm 0cm 3cm, width=0.85\textwidth]{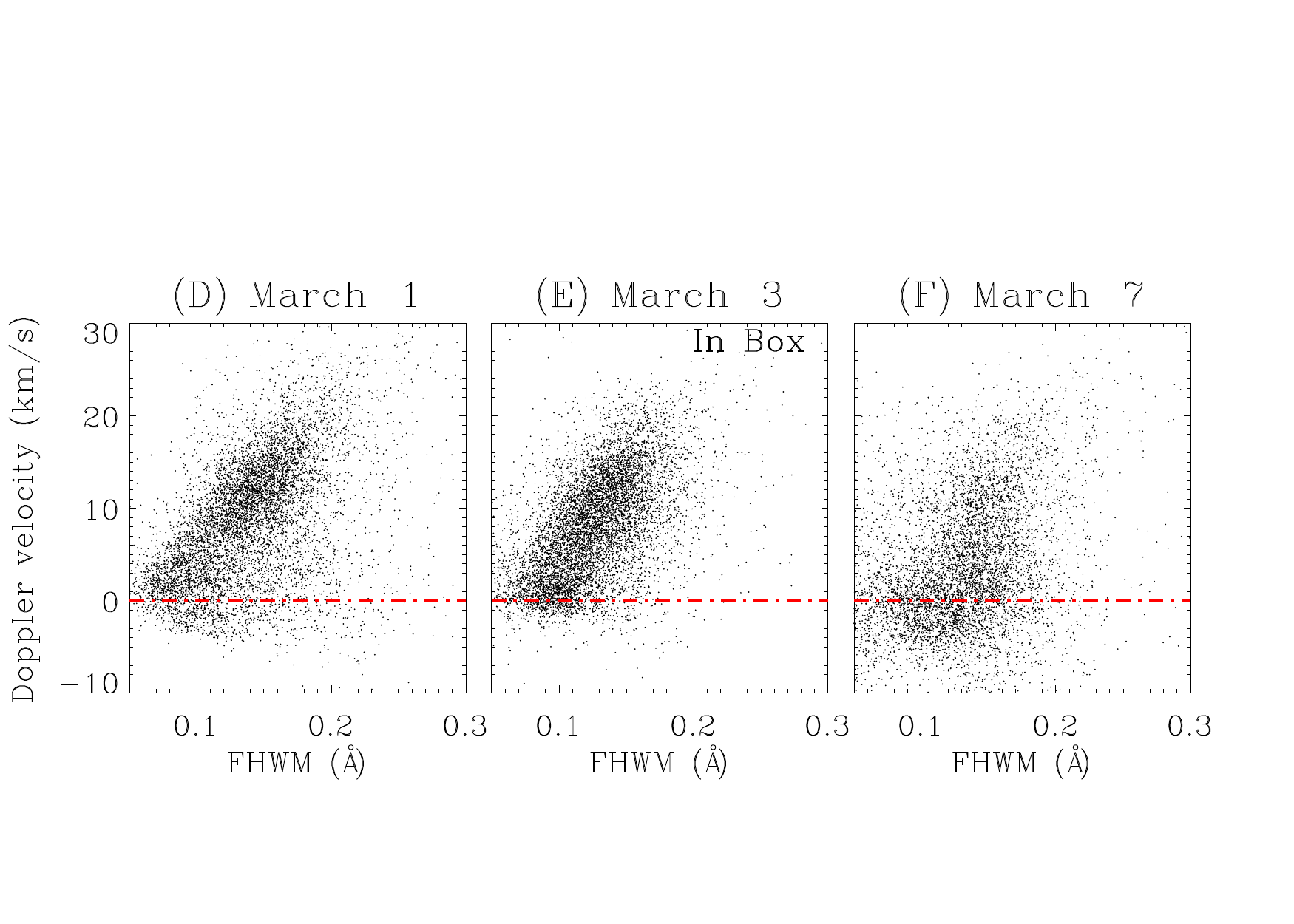}
    \caption{Top row: Scatter plot of Doppler velocity versus FWHM obtained using \ion{Si}{4}~1393.78~{\AA} for the strong field regions. Bottom row: Same for the section of the weak field corridor contained within the box region, shown in Figs.~\ref{ntv1} \& \ref{ntv2}. The red horizontal lines in each panel show the zero velocity mark.} \label{scatter_dv_wd}
\end{figure}
\begin{figure}[t!] 
\centering
\includegraphics[trim=0.cm 0.cm 0.cm 0.cm,width= 1.0\textwidth]{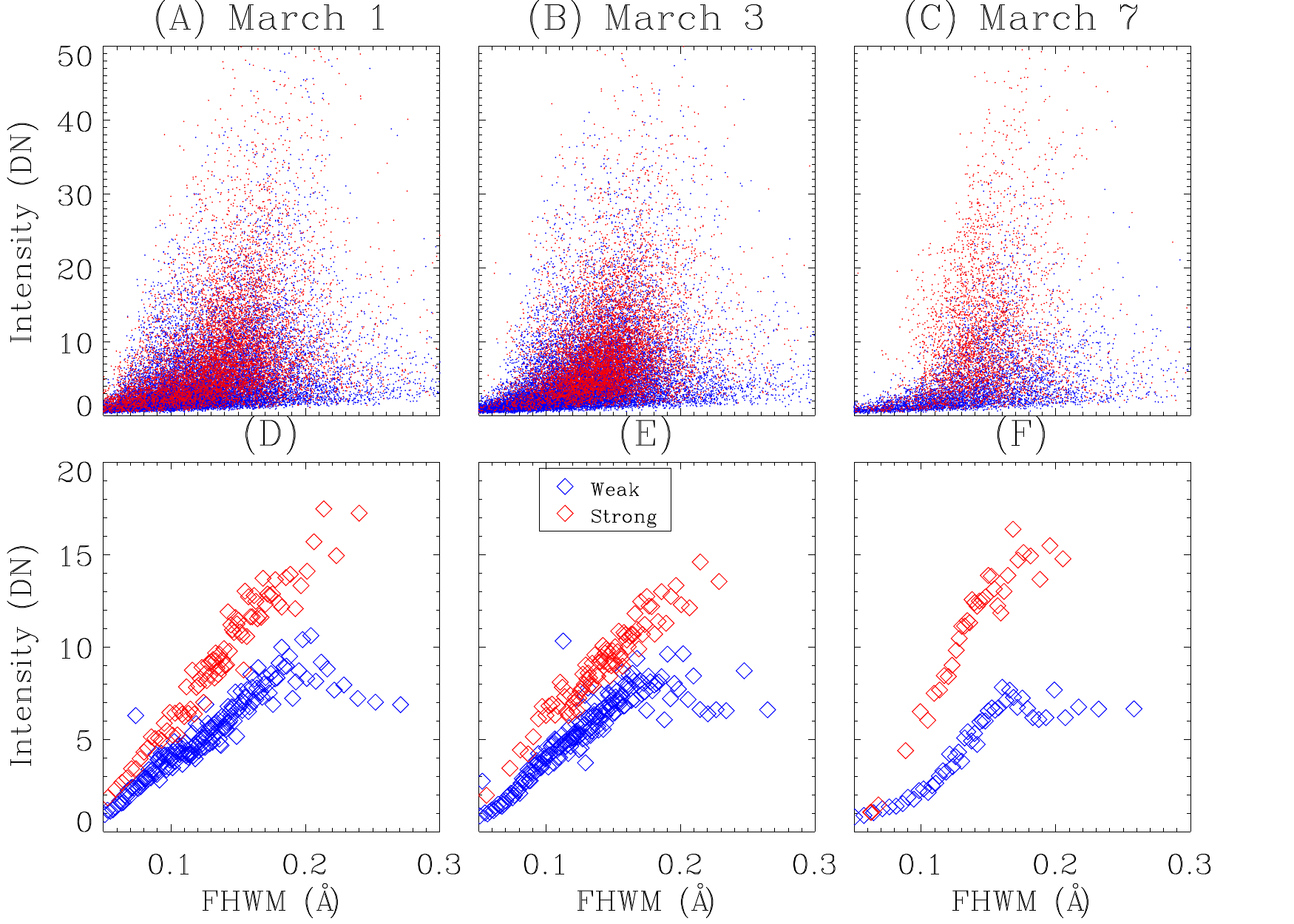}
\caption{Top panels: Scatter plot of intensity versus FWHM obtained using \ion{Si}{4}~393.78~{\AA} line for strong (red) and weak field (blue) regions (threshold is 50~G). Bottom panels: Same as above but the scatter being binned in sets of 100 points. The instrumental width for IRIS $i.e.,$ 0.02~{\AA} has been subtracted from the FWHM values.\color{black} }\label{scat2}
\end{figure}
\clearpage
\thispagestyle{empty}
\centering
\vspace{1cm}
\captionof{table}{Table with detailed information regarding the dates of observations (March 1{--}8, 2017) and the $\mu$-value of the centre of the FOV in the first and second columns, respectively. The third (fifth) column shows the mean non-thermal velocities along with their standard errors (see Appendix in Paper-I for details) in the strong field regions (corridors). The corridors are defined by the white boxes in Figs.~\ref{ntv1} \& \ref{ntv2}. The fourth (sixth) column shows the median values of the non-thermal velocities for the strong field (corridor) regions. The rasters on March 6 have a lot of missing exposures and those on March 7{--}8 are very close to the limb, hence the results could not be realistic estimates.}\label{tab_ntv}
\label{details}
\vspace{1cm}
 \begin{tabular}{| c | c | c | c | c | c |}  \hline
Date     &$\mu$  & \multicolumn{2}{c|}{Strong field}   & \multicolumn{2}{c|}{Corridor}  \\
 2017       & &\multicolumn{2}{c|}{(km s$^{-1}$)}   &\multicolumn{2}{c|}{(km s$^{-1}$)} \\
         & &\multicolumn{2}{c|}{\ion{Si}{4}} &\multicolumn{2}{c|}{\ion{Si}{4}} \\ \hline
  &    &Mean  &Median  &Mean &Median  \\ \hline 
1 March   &0.82  &18.2$\pm$0.06 &18.7 &18.4$\pm$0.10 &18.6 \\
2 March  &0.90  &18.9$\pm$0.05 &19.2 &16.7$\pm$0.10 &15.9 \\
3 March  &0.92  &19.5$\pm$0.05 &19.7 &16.5$\pm$0.09 &16.5 \\
4 March  &0.89  &19.6$\pm$0.05 &19.6 &16.8$\pm$0.09 &16.7 \\
5 March  &0.81 &19.5$\pm$0.06 &19.7 &17.4$\pm$0.10 & 17.4 \\
6 March  &0.68  &18.8$\pm$0.07 &18.9  &16.2$\pm$0.10 & 16.4 \\
7 March  &0.51  &19.6$\pm$0.07 &19.6 &16.9$\pm$0.11 &17.4 \\ 
8 March  &0.27 &20.3$\pm$0.09 &20.5 &20.0$\pm$0.10 &20.2 \\ \hline
\end{tabular}
\end{document}